\documentclass[prd, twocolumn, nofootinbib, showpacs, amsmath, amssymb, floatfix, eqsecnum]{revtex4-1}
\usepackage{amsmath}
\usepackage{amssymb}
\usepackage{amsthm}
\usepackage{amsfonts}

\usepackage{graphicx,color,framed}
\usepackage{hyperref}
\usepackage{times}
\usepackage{enumerate}
\usepackage{lipsum}
\usepackage{slashed}
\usepackage{url}
\usepackage{bbm}
\hypersetup{
    colorlinks=true, 
    linktoc=all,     
    linkcolor=blue,  
}

\def \beq {\begin{equation}}
\def \eeq {\end{equation}}
\def \beqa {\begin{eqnarray}}
\def \eeqa {\end{eqnarray}}
\def \bseq {\begin{subequations}}
\def \eseq {\end{subequations}}

\newtheorem{theorem}{Theorem}

\begin{document}

\title{Note on quantum cellular automata and strong equivalence}

\author{Carolyn Zhang}
\affiliation{Department of Physics, Kadanoff Center for Theoretical Physics, University of Chicago, Chicago, Illinois 60637,  USA}
\date{\today}

\begin{abstract}
In this note, we present some results on the classification of quantum cellular automata (QCA) in 1D under strong equivalence rather than stable equivalence. Under strong equivalence, we only allow adding ancillas carrying the original on-site representation of the symmetry, while under stable equivalence, we allow adding ancillas carrying any representation of the symmetry. The former may be more realistic, because in physical systems especially in AMO/quantum computing contexts, we would not expect additional spins carrying arbitrary representations of the symmetry to be present. Ref.~\onlinecite{mpu} proposed two kinds of symmetry-protected indices (SPIs) for QCA with discrete symmetries under strong equivalence. In this note, we show that the more refined of these SPIs still only has a one-to-one correspondence to equivalence classes of $\mathbb{Z}_N$ symmetric QCA when $N$ is prime. We show a counter-example for $N=4$. We show that QCA with $\mathbb{Z}_2$ symmetry under strong equivalence, for a given on-site representation, are classified by $\mathbb{Z}^{pq}$ where $p$ is the number of prime factors of the on-site Hilbert space dimension and $q$ is the number of prime factors of the trace of the nontrivial on-site $\mathbb{Z}_2$ element. Finally, we show that the GNVW index has a formulation in terms of a $\mathbb{Z}_2$ SPI in a doubled system, and we provide a direct connection between the SPI formulation of the GNVW index and a second Renyi version of the mutual information formula for the GNVW index.
\end{abstract}

\maketitle

\section{Introduction}\label{sintroduction}

Some unitary operators that preserve locality, mapping local operators to nearby (strictly local) operators, cannot be written as finite-depth quantum circuit (FDQC). Such strict locality-preserving unitaries, also known as quantum cellular automata (QCA), are peculiar because their actions cannot be spatially restricted. QCA have been studied and classified in low dimensions\cite{GNVW,fermionic, freedmanclassification,freedmangroup, haahnontrivial, haahclifford, moreqca,mpusymm, mpu, pirolifermionic, xiempu, piroliqca, u1floquet}. In 1D, a simple example of a nontrivial QCA is a translation operator. A translation operator preserves locality, mapping an operator $O_i$ on site $i$ to $O_{i+1}$ on site $i+1$. However, it cannot be generated by any local Hamiltonian, and therefore cannot be restricted. In fact, all nontrivial QCA in 1D are generalized translations, and can be classified using a ``GNVW index"\cite{GNVW, chiralbosons, fermionic, mpu, u1floquet, zhang2022} that takes value in the rational numbers\cite{GNVW,chiralbosons,fermionic,mpu,harperorder}. In the presence of $U(1)$ symmetry, 1D QCA are completely classified by a ratio of two functions $\pi(z)=\frac{a(z)}{b(z)}$\cite{u1floquet}. For discrete symmetries, the classification depends on the definition of equivalence. As we will discuss in more detail below, \emph{stable equivalence} leads to a classification given by $H^2(G,U(1))$ (in addition to translations). On the other hand, \emph{strong equivalence} leads to a richer classification. Ref.~\onlinecite{mpu} studied this classification and suggested a set of symmetry protected indices (SPIs), one for each symmetry group element. Roughly speaking, the difference between stable equivalence and strict equivalence is that for the former, we allow tensoring with ancillas carrying any finite dimensional representation of the symmetry, while for the latter we only allow tensoring with ancillas carrying the original on-site representation of the symmetry. Ref.~\onlinecite{mpu} also provided a way to measure these SPIs in interferometric experiments.

For QCA with $U(1)$ symmetry, the physical meaning of the charge part of $\pi(z)$, given by $\tilde{\pi}(z)$, is clear: it corresponds to the net $U(1)$ charge transported in the positive direction upon application of the QCA. The index can be therefore detected by measuring charge transport\cite{u1floquet, flows}. The GNVW index has been interpreted as net transport of quantum information\cite{chiralbosons}, but it is less obvious how to measure this transport. Ref.~\onlinecite{tracking} (see also Ref.~\onlinecite{ranard2022}) showed that this index can be measured using quantum mutual information if we consider a setting where a subset of the spins in the spin chains are initially maximally entangled with a partner ancilla. Specifically, the mutual information quantity measures the change in entanglement across a cut, due to the entanglement between the spins and the ancillas. While this gives some physical intuition for the GNVW index, their mutual information formula was not directly derived from the original GNVW index formula. 

In this note, we present the following collection of results related to the classification of QCA with discrete symmetries, under strong equivalence:

\begin{itemize}
\item{A proof that the refined SPI of Ref.~\onlinecite{mpu} completely classifies $\mathbb{Z}_N$ symmetric QCA under strong equivalence when $N$ is prime, and a counterexample for $N=4$.}
\item{A derivation that $\mathbb{Z}_2$ symmetric QCA are classified by $\mathbb{Z}^{pq}$ where $p,q$ are the number of prime factors of the on-site Hilbert space dimension and trace of the nontrivial on-site $\mathbb{Z}_2$ element respectively.}
\item{A connection between the GNVW index and the $\mathbb{Z}_2$ SPI of a doubled system.}
\item{A direct connection betwen the SPI formulation of the GNVW index and a second Renyi version of the mutual information formula in Ref.~\onlinecite{tracking} for the GNVW index, when the ancillas are initially maximally entangled with the edge spins. Note that there is an alternate derivation of this result using tensor network methods in the Supplemental material of Ref.~\onlinecite{gong2021chaos}.}
\end{itemize}
\section{Setting and definitions}
We consider a periodic chain of bosonic $d$-state spins in 1D evolving under a QCA $U$, that respects a global unitary symmetry $G$. For simplicity, we will only consider $G=\mathbb{Z}_N$ in this work. In this context, we will proceed with some definitions that will be useful for specifying the problems we would like to solve.

\subsection{QCA and strong equivalence}
We define a QCA as a unitary operator that takes any operator fully supported on site $i$ to an operator supported strictly on $[i-\xi,i+\xi]$, where $\xi$ is a bounded operator spreading length. We define two 1D QCA $U$ and $U'$ as equivalent if they differ by a 1D FDQC $W$:
\begin{equation}
U'=W\cdot U.
\end{equation}

Without loss of generality (for 1D), we can assume that $W$ consists of two layers of local unitaries, where in each layer the unitaries are all supported on disjoint, bounded intervals. We say that $U$ is $G$ symmetric if it commutes with the global symmetry operators $U_g$ for all $g\in G$:
\begin{equation}
[U,U_g]=0\qquad\forall g\in G.
\end{equation}

Here, $U_g=\prod_{i\in\Lambda}\mu_{g,i}$ is the unitary representation of $G$ for the entire system and $\mu_{g,i}$ is the on-site representation of $G$. We assume that the on-site representation is identical on each site. In the presence of symmetry, $U'\sim U$ if they differ by a $G$ symmetric FDQC, which is an FDQC built out of gates that each individually commute with $U_g$ for all $g\in G$.

Equivalence relations in the presence of symmetries are usually defined using \emph{stable equivalence}, which allows for the addition of ancillas carrying arbitrary representations of $G$ to be added to each site. Ancillas are degrees of freedom that can transform nontrivially under the symmetry but evolve trivially under $U$, so the entire system is described by $U_{\mathrm{tot}}=U\otimes\mathbbm{1}_a$, where $\mathbbm{1}_a$ acts on the ancillas. Specifically, under stable equivalence, $U\sim U'$ if and only if $U\otimes\mathbbm{1}_{a}=W\cdot (U'\otimes\mathbbm{1}_{a'})$ where $W$ is a $G$ symmetric FDQC and $\mathbbm{1}_{a}$ and $\mathbbm{1}_{a'}$ are act on two different sets of ancillas. Under stable equivalence, 1D QCA with symmetry have been classified by cohomology\cite{cohomology,alldimensions,dynamically}. The classification is given by the GNVW index\cite{GNVW,chiralbosons} along with an element of $H^2(G,U(1))$.

We may also study the case where we only allow ancillas with the same on-site representation of $G$ as the original sites. As mentioned earlier, this representation is given by $\mu_g$. This definition of equivalence may be considered more natural because in a real physical system, one would not expect additional spins with arbitrary representations of $G$ to be present. In Ref.~\onlinecite{mpu}, this definition of equivalence was called \emph{strong equivalence}. This is the definition of equivalence we will use in the rest of this note, unless otherwise specified.

For systems without symmetry and systems with $U(1)$ symmetry, we do not expect there to be a difference between the stable equivalence classification and the strong equivalence classification. However, for systems with discrete symmetries, aspects of the strong equivalence classification are still not understood. 

\subsection{GNVW index}
In the absence of symmetry, QCA in 1D are classified by the GNVW index\cite{GNVW}. Here, we will review the GNVW index because it will come up frequently later in this note. 

The GNVW index measures net translation of operators in the spin chain. Specifically, to each site in the spin chain, we can associate an operator algebra, which is the algebra of operators fully supported on that site. Using the tensor product structure of the Hilbert space, we can then obtain the operator algebra for an interval $A$ of sites. We denote this algebra by $\mathcal{A}$. We define an overlap function $\eta(\mathcal{A},\mathcal{B})$ of two operator algebras $\mathcal{A}$ and $\mathcal{B}$ as
\begin{equation}\label{etaoperators}
\eta(\mathcal{A},\mathcal{B})=\sqrt{\sum_{\substack{O_a\in\mathcal{A} \\ O_b\in\mathcal{B}}}|\mathrm{tr}(O_a^\dagger O_b)|^2},
\end{equation}
where the sum runs over an orthonormal basis of operators in $\mathcal{A}, \mathcal{B}$. In (\ref{etaoperators}), the lowercase symbol ``$\mathrm{tr}$'' denotes a normalized trace defined by $\mathrm{tr}(\mathbbm{1}) = 1$. $\eta(\mathcal{A},\mathcal{B})$ simply counts the number of operators contained in both $\mathcal{A}$ and $\mathcal{B}$. It is easy to see that $\eta(\mathcal{A},\mathcal{A})=d^{|A|}$ and $\eta(\mathcal{A},\mathcal{B})=1$ if $A\cap B=\emptyset$. Using this overlap function, the GNVW index is given by\cite{GNVW}\footnote{The index is sometimes defined as the logarithm of (\ref{gnvw}).}
\begin{equation}\label{gnvw}
\mathrm{ind}(U)=\frac{\eta(U^\dagger\mathcal{A}U,\mathcal{B})}{\eta(\mathcal{A},U^\dagger \mathcal{B}U)},
\end{equation}
where $\mathcal{A}, \mathcal{B}$ are operator algebras of two adjacent intervals $A$ and $B$. The only constraint on $A$ and $B$ is that they must be at least twice the operator spreading length of $U$. The intuition behind $\mathrm{ind}(U)$ is that it gives the change in the overlap of the operator algebras $\mathcal{A}$ and $\mathcal{B}$ due to the net chiral action of $U$. For example, if $U$ is a translation by a single lattice site, then $\mathrm{ind}(U)=d$.

\subsection{Symmetry-protected indices}
To classify $G$ symmetric QCA, Ref.~\onlinecite{mpu} proposed using a set of symmetry-protected indices (SPIs), consisting of an index for each group element. The SPI for a group element $g$ can be nontrivial if $\mathrm{Tr}(\mu_g)\neq 0$, where $\mu_g$ is the on-site representation of the symmetry. To define the SPIs, we use the action of $U$ on the global symmetry operator $U_g$ restricted to an interval $I$ with an even number of sites. Because $U_g$ is a product of on-site operators, this restriction is unambiguous. We denote the restriction of $U_g$ to $I$ by
\begin{equation}\label{ugi}
U_{g,I}=U_{g,L}\otimes U_{g,R},
\end{equation}
where $U_{g,L}$ and $U_{g,R}$ are supported on the left and right halves of $I$ respectively. The action of $U$ on $U_{g,I}$ is given by
\begin{equation}\label{rgdef}
U^\dagger U_{g,I}U=L_g\otimes R_g,
\end{equation}
where $R_g$ and $L_g$ are supported to the right and the left of the midpoint of $I$ respectively. Although $U_{g,I}\sim L_g\otimes R_g$ as representations (meaning their characters match on all group elements), $U_{g,R}$ and $R_g$ may not be equivalent as representations. Notice that there is an important ambiguity in defining $R_g$ and $L_g$, because $R_g$ can carry a 1D representation of $G$ that is canceled by $L_g$. We prove in Appendix~\ref{sderivation} the following important theorem: 
\begin{theorem}\label{th1}
Two $G$ symmetric QCA $U$ and $U'$ are equivalent if and only if their corresponding representations $R_g$ and $R_g'$ as defined in (\ref{rgdef}) differ only by a 1D representation $u_g$ of $G$ and tensoring with on-site representations $\mu_g$.
\end{theorem} 
In particular, $R_g=u_g\mu_g^{\otimes N}$ for some nonnegative integer $N$ if and only if $U$ is a $G$ symmetric FDQC. 

Note that this result was derived using a matrix product state formalism in Ref.~\onlinecite{mpu}. However, the problem of finding concrete topological invariants for $U$ remains unsolved. Theorem~\ref{th1} indicates that we should seek quantities that are insensitive to $R_g\to u_gR_g$ where $u_g$ is a 1D representation of $G$. The symmetry-protected indices (SPIs), defined in Ref.~\onlinecite{mpu} were designed to be insensitive to $u_g$. These SPIs, denoted by $\{\mathrm{ind}_g\}$, are given by\footnote{Ref.~\onlinecite{mpu} uses the logarithm of the indices presented here, which are additive rather than multiplicative under stacking.}
\begin{equation}
\mathrm{ind}_g(U)=\mathrm{ind}(U)\cdot\sqrt{\left|\frac{\mathrm{Tr}(R_g)}{\mathrm{Tr}(L_g)}\right|},
\end{equation}
where $\mathrm{ind}(U)$ is the GNVW index of $U$ and the trace in the numerator and denominator are all taken over the same space dimension (so $\mathrm{Tr}(R_0)=\mathrm{Tr}(L_0)$, where $0$ is the trivial group element). Note that $\mathrm{ind}_0(U)=\mathrm{ind}(U)$.

While $\{\mathrm{ind}_g(U)\}$ are certainly invariant for QCA that are equivalent, they are \emph{not} complete. In other words, they can match for two QCA that do not differ by a $G$ symmetric FDQC. A simple example given in the supplemental material of Ref.~\onlinecite{mpu} uses $G=\mathbb{Z}_3$ and an on-site representation $\mu_g=\mathrm{diag}(1,e^{2\pi i/3},e^{2\pi i/3})$. In this case, we can have $R_g=L_g=\mathrm{diag}(1,e^{4\pi i/3},e^{4\pi i/3})$ because $L_g\otimes R_g=\mu_g\otimes \mu_g$. Clearly, $\mathrm{ind}_g=1$ for all $g\in G$, even though $R_g$ is not equivalent to $U_{g,R}=\mu_g$.

\subsection{Refined symmetry-protected indices}
To remedy the fact that $\{\mathrm{ind}_g(U)\}$ are not complete, Ref~\onlinecite{mpu} also proposed a set of refined SPIs (see Eq. 89 in the supplemental material). These refined indices are defined as
\begin{equation}\label{rinddef}
\mathrm{rind}_g(U)=\mathrm{ind}(U)\cdot\left[\frac{\mathrm{Tr}(R_g)}{\mathrm{Tr}(U_{g,R})}\right]^{d_g},
\end{equation}
where $d_g$ is the order of the group element $g\in G$. Again, the trace in the numerator and that in the denominator are taken over the same Hilbert space, so $\mathrm{rind}_1(U)=\mathrm{ind}(U)$. The purpose of taking the $d_g$ power is to remove the ambiguity of the 1D representation $u_g$. For the $\mathbb{Z}_3$ example above, it is easy to see that $\mathrm{rind}_g=-1\neq 1$ for each of the two nontrivial elements of $\mathbb{Z}_3$.

\section{Limitations of refined symmetry-protected indices}\label{slim}
With the setting and previous results established, we will now introduce some new results. First, we will show that the refined SPIs $\{\mathrm{rind}_g(U)\}$ only completely classify $\mathbb{Z}_N$ symmetric QCA when $N$ is prime, and give a counterexample for $N=4$. 

In order for the refined SPIs to completely classify $\mathbb{Z}_N$ symmetric QCA, it must be true that
\begin{equation}\label{iff}
\mathrm{Tr}(R_g)^{d_g}=\mathrm{Tr}(R_g')^{d_g}\qquad\mathrm{iff}\qquad R_g=u_gR_g',
\end{equation}
where $R_g$ and $R_g'$ are representations of the symmetry and $u_g$ is a 1D representation of the symmetry. The ``if" direction follows immediately. For the ``only if" direction, let us write
\begin{equation}
R_g=\oplus_ra_ru_{r,g}\qquad R_g'=\oplus_rb_ru_{r,g}.
\end{equation}

Here, $a_r$ is the number of 1D irreducible representations $e^{2\pi ir/N}$ of $\mathbb{Z}_N$ in $R_g$, and $u_{r,g}=\left(u_r\right)^g$. Let us assume that $\mathrm{Tr}(R_g)^{d_g}=\mathrm{Tr}(R_g')^{d_g}$. For (\ref{iff}) to be true, we must have $a_r=b_{r-r'}$ where $r'$ depends on the 1D representation $u_g$. 

We will now evaluate $\mathrm{Tr}(R_0)^1$ and $\mathrm{Tr}(R_1)^N$ and compare them to $\mathrm{Tr}(R_0')^1$ and $\mathrm{Tr}(R_1')^N$. Here, $R_1$ is the generator of $\mathbb{Z}_N$. Setting $\mathrm{Tr}(R_0)^1=\mathrm{Tr}(R_0')^1$, we get
\begin{equation}\label{g0}
\sum_ra_r=\sum_rb_r.
\end{equation}

Next, setting $\mathrm{Tr}(R_1)^N=\mathrm{Tr}(R_1')^N$, we get
\begin{equation}
\sum_ra_ru_{r,1}=v_1\sum_rb_ru_{r,1},
\end{equation}
where $v_1^N=1$. Suppose that $v_1=u_{r',1}$. Then $v_1$ shifts $u_{r,1}\to u_{r+r',1}$, so
\begin{equation}
\sum_ru_{r,1}(a_r-b_{r-r'})=0.
\end{equation}

For prime $N$, only the sum of all the different 1D reps give zero (i.e. $\sum_ru_{r,1}=0$), so $a_r-b_{r-r'}$ must be the same nonnegative integer $n\in\mathbb{N}$ for all $r$. But if $a_r-b_{r-r'}=n$ for all $r$ and $n\neq 0$, then (\ref{g0}) is not satisfied. So $a_r=b_{r-r'}$ for all $r$.

This argument fails for $N$ not prime. For example if we take $N=4$, there are two ways to get 0, either by having equal numbers of $1$ and $-1$ or equal numbers of $i$ and $-i$. One can then construct representations for which $\mathrm{Tr}(R_g)^{d_g}=\mathrm{Tr}(R_g')^{d_g}$ for all group elements $g$, but $R_1\neq u_1R_1'$. 

One example for $N=4$ is $R_1=\mathrm{diag}(1,1,-1,i,-i)$ and $R_1'=\mathrm{diag}(1,i,i,-i,-i)$ (here we added an extra trivial rep so that the character is not zero for any group element). One can check that the characters for $g=(0,1,2,3)$ are $(5,1,3,1)$ for $R_g$ and $(5,1,-3,1)$ for $R_g'$, so they are different representations, differing by more than just a 1D rep $u_g$. However, their characters to the $d_g$th power are equal for all $g$.

In summary, we find that even the refined indices are not fine enough to completely classify QCA with $\mathbb{Z}_N$ symmetry under strong equivalence. They only completely classify $\mathbb{Z}_N$ symmetric QCA when $N$ is prime.

\section{Classification and invariants of $\mathbb{Z}_2$-symmetric QCA}\label{sclass}
Despite the known abstract statement of the classification given by Theorem~\ref{th1}, concrete topological invariants for $G$ symmetric QCA are difficult to find, as is clear from the discussion above. In this section, we will study the simple case where $G=\mathbb{Z}_2$, for which we can remove the ambiguity of the 1D representation $u_g$ by simply taking an absolute value. We then completely classify QCA with $\mathbb{Z}_2$ symmetry, by finding all allowed values for the $\mathbb{Z}_2$ SPI, given an on-site representation $\mu_g$ of $\mathbb{Z}_2$. We will then show that (1) the GNVW index can be interpreted as a $\mathbb{Z}_2$ SPI in a doubled system and (2) the $\mathbb{Z}_2$ SPI links the original GNVW index and a formulation of it in terms of mutual information, given in Ref.~\onlinecite{tracking,ranard2022}.

\subsection{Classification of $\mathbb{Z}_2$ symmetric QCA}\label{sz2}
In this section, we work out the complete classification of 1D QCA with $\mathbb{Z}_2$ symmetry under strong equivalence, given an on-site representation $\mu_g$ of the $\mathbb{Z}_2$ symmetry. 

The only irreducible representations of $\mathbb{Z}_2$ is the trivial representation $(1)$ and the sign representation $(-1)$. Therefore, we can remove the ambiguity of the 1D representation of $G$ attached to $R_1$, where $g=1$ is the generator of $\mathbb{Z}_2$, by taking the absolute value of $\mathrm{Tr}(R_1)$. We define
\begin{equation}\label{indices}
\pi_0(U)=\mathrm{ind}(U)\qquad \pi_{1}(U)=\pi_0(U)\bigg|\frac{\mathrm{Tr}(R_1)}{\mathrm{Tr}(U_{1,R})}\bigg|,
\end{equation}
where, as always, the trace in the numerator and denominator are over the same Hilbert space. We will refer to $\pi_1(U)$ as the $\mathbb{Z}_2$ SPI from this point onward.

Suppose that the spectrum of $\mu_1$ has the eigenvalue $+1$ with multiplicity $a_0$ and the eigenvalue $-1$ with multiplicity $a_1$. Then the representation of $G$ given by $\mu_g$ is completely defined by

\begin{align}
\begin{split}
\mathrm{Tr}(\mu_0)&=d=a_0+a_1\\
\mathrm{Tr}(\mu_1)&=\chi=a_0-a_1.
\end{split}
\end{align}

We now claim that the complete classificaton of QCA with $\mathbb{Z}_2$ symmetry is given by all the vectors $(\pi_0(U),\pi_1(U))$ with components $\pi_0(U)$ and $\pi_1(U)$ satisfying
\begin{align}
\begin{split}\label{conditions}
d^{N_1}\pi_{0}(U)&=\alpha_{0}+\alpha_{1}\\
\frac{d^{N_2}}{\pi_{0}(U)}&=\beta_{0}+\beta_{1}\\
\chi^{N_1}\pi_{1}(U)&=\alpha_{0}-\alpha_{1}\\
\frac{\chi^{N_2}}{\pi_{1}(U)}&=\beta_{0}-\beta_{1},
\end{split}
\end{align}
where $N_1,N_2,\alpha_{0},\alpha_{1},\beta_0,$ and $\beta_{1}$ are all nonnegative integers. We will first provide the classification obtained from (\ref{conditions}), and then we will justify (\ref{conditions}).
\subsubsection{Classification result}\label{sclassresult}

If $\chi=0$, then (\ref{conditions}) says that
\begin{equation}
\pi_1(U)=\frac{\alpha_0-\alpha_1}{\chi^{N_1}}=\frac{\chi^{N_2}}{\beta_0-\beta_1}.
\end{equation}

Therefore, for $\pi_1(U)$ to be defined, it must be 1. In this case, all that remains is the no-symmetry GNVW classification, in agreement with Ref.~\onlinecite{mpu}. This classification is simply given by $\mathbb{Z}^p$ where $p$ is the number of prime factors of $d$. Notice that $\pi_1(U)=1$ means that $\alpha_0=\alpha_1$ and $\beta_0=\beta_1$ so $\alpha_0+\alpha_1$ and $\beta_0+\beta_1$ is always even. However, since $d$ is also even for $G=\mathbb{Z}_2$, $d^{N_1}\pi_0(U)$ and $\frac{d^{N_2}}{\pi_0(U)}$ in (\ref{conditions}) are also even for sufficiently large $N_1$ and $N_2$, even if $\pi_0(U)$ is odd. 

Suppose that $\chi\neq 0$, giving extra constraints on the allowed $\pi_1(U)$. For simplicity first consider the case where $\pi_0(U)=1$ so these phases are only nontrivial in the presence of $\mathbb{Z}_2$ symmetry. Since $H^2(\mathbb{Z}_2,U(1))=\mathbb{Z}_1$, the phases studied here are in fact only nontrivial under strong equivalence. Eq.~\ref{conditions} simplifies to
\begin{align}
\begin{split}
\alpha_0&=\frac{\chi^{N_1}\pi_{1}(U)+d^{N_1}}{2}\qquad \alpha_1=d^{N_1}-\alpha_0\\
\beta_0&=\frac{\chi^{N_2}/\pi_{1}(U)+d^{N_2}}{2}\qquad \beta_1=d^{N_2}-\beta_0.
\end{split}
\end{align}

It is not hard to see that for any $\pi_{1}(U)$ with the same prime factors as $\chi$, one can always find sufficiently large $N_1$ and $N_2$ that gives nonnegative integer values for $\alpha_{0},\alpha_{1},\beta_0,$ and $\beta_{1}$. In particular, notice that if $d$ is odd, then $\chi$ and $\pi_{1}(U)$ must also be odd, so $\chi^{N_1}\pi_{1}(U)+d^{N_1}$ and $\chi^{N_2}/\pi_{1}(U)+d^{N_2}$ are both divisible by two. On the other hand if $d$ is even, then $\chi$ must also be even, so $\chi^{N_1}\pi_{1}(U)+d^{N_1}$ and $\chi^{N_2}/\pi_{1}(U)+d^{N_2}$ is also divisible by two for sufficiently large $N_1,N_2$. The enrichment of the classification due to $\mathbb{Z}_2$ symmetry is therefore $\mathbb{Z}^{q}$ where $q$ is the number of prime factors of $\chi$. Combining this with the classification in the absence of $\mathbb{Z}_2$ symmetry, we obtain a total classification of $\mathbb{Z}^{pq}$.

\subsubsection{Justification of (\ref{conditions})}\label{sjust}
We now show that Eq.~\ref{conditions} correspond to all the distinct equivalence classes, labeled by $(\pi_0(U),\pi_1(U))$, realizable in the system. In other words, we will prove (1) if $U$ is a $\mathbb{Z}_2$ symmetric QCA, then (\ref{conditions}) are satisfied and (2) if (\ref{conditions}) are satisfied, then we can construct a corresponding $\mathbb{Z}_2$ symmetric QCA.

For (1), note that if $U$ is a $\mathbb{Z}_2$ symmetric QCA, then
\begin{equation}
U^\dagger U_g U=U_g
\end{equation}
This means that
\begin{equation}\label{speccon}
\mathrm{Spec}\left(U_{g,L}U_{g,R}\right)=\mathrm{Spec}\left(L_gR_g\right).
\end{equation}
Let $|I|=N_1+N_2$. Then (\ref{conditions}) means that
\begin{align}
\begin{split}\label{alphabeta}
d^{N_1+N_2}&=\left(\alpha_0+\alpha_1\right)\left(\beta_0+\beta_1\right)\\
\chi^{N_1+N_2}&=\left(\alpha_0-\alpha_1\right)\left(\beta_0-\beta_1\right).
\end{split}
\end{align}

Identifying $\alpha_0$ and $\beta_0$ with the number of $+1$ eigenvalues of $L_g$ and $R_g$ respectively, and $\alpha_1$ and $\beta_1$ with the number of $-1$ eigenvalues of $L_g$ and $R_g$, we see that (\ref{alphabeta}) follows from (\ref{speccon}), with nonnegative $\alpha_0,\alpha_1,\beta_0,$ and $\beta_1$. Therefore, (\ref{speccon}) implies (\ref{conditions}).

For (2), we will construct a $\mathbb{Z}_2$ symmetric QCA given $\alpha_0,\alpha_1,\beta_0,$ and $\beta_1$. To begin, we cluster groups of $2N_1+N_2$ sites into supersites, with Hilbert space 
\begin{equation}
\mathcal{H}^{2N_1+N_2}=\mathcal{H}^{N_1}\otimes\mathcal{H}_{\alpha}\otimes\mathcal{H}_{\beta}
\end{equation}
where the representation of the $\mathbb{Z}_2$ symmetry is given by $\mu_g^{N_1}$ on $\mathcal{H}^{N_1}$, and $L_g$ and $R_g$ on $\mathcal{H}_{\alpha}$ and $\mathcal{H}_{\beta}$ respectively. Now we consider a QCA that performs a unit supersite translation on the $\mathcal{H}_{\alpha}$ sites, and a unit supersite translation on the $\mathcal{H}^{N_1}$ sites in the opposite direction, and does nothing to the $\mathcal{H}_{\beta}$ sites. This would give
\begin{equation}
\pi_0(U)=\frac{\alpha_0+\alpha_1}{d^{N_1}}\qquad\pi_1(U)=\frac{\alpha_0-\alpha_1}{\chi^{N_1}},
\end{equation}
as desired.

\subsection{The GNVW index in terms of a $\mathbb{Z}_2$ SPI}\label{sphysical}

In this section, we show that the GNVW index $\mathrm{ind}(U)$ can be interpreted in terms of a $\mathbb{Z}_2$ SPI for the $\mathbb{Z}_2$ SWAP symmetry of $U\otimes U$ acting in a doubled system. We begin with a useful formulation of the overlap function $\eta(\mathcal{A},\mathcal{B})$ derived in Ref.~\onlinecite{flows}. There, it was shown that in a doubled system with $2|\Lambda|$ total sites, we can write the overlap function in terms of a SWAP operator between the two spin chains. By SWAP operator, we simply mean an operator with the following action:
\begin{equation}
\mathrm{SWAP}_{i,j}^\dagger \left(O_i\otimes O_j'\right)\mathrm{SWAP}_{i,j}=\left(O_i'\otimes O_j\right),
\end{equation}
where $O_i$ is an operator supported on site $i$ and $O_j$ is the same operator translated to site $j$. Similarly, $O_j'$ is supported on site $j$ and $O_i'$ is the same operator, but translated to site $i$. Any SWAP operator is unitary and order two, and hence is also Hermitian. Denoting the two spin chains by $\Lambda_1$ and $\Lambda_2$, with $A_1,B_1\subset\Lambda_1$ and $A_2,B_2\subset\Lambda_2$, we have\cite{flows}
\begin{align}
\begin{split}\label{swapoverlap}
\eta&(\mathcal{A},\mathcal{B})=\frac{d^{(N_A+N_B)/2}}{d^N}\\
&\times\sqrt{\mathrm{Tr}\left(\mathrm{SWAP}_{A_1,A_2}\mathrm{SWAP}_{B_1,B_2}\right)}.
\end{split}
\end{align}

To obtain $\eta(U^\dagger\mathcal{A}U,\mathcal{B})$, we define an operator $U_{1,2}=U\otimes U$ that acts as the QCA $U$ on both $\Lambda_1$ and $\Lambda_2$:
\begin{align}
\begin{split}\label{swapindex}
\eta&(U^\dagger\mathcal{A}U,\mathcal{B})=\frac{d^{(N_A+N_B)/2}}{d^N}\\
&\times\sqrt{\mathrm{Tr}\left(U_{1,2}\mathrm{SWAP}_{A_1,A_2}U_{1,2}^\dagger\mathrm{SWAP}_{B_1,B_2}\right)}.
\end{split}
\end{align}

Since $\mathrm{SWAP}_{A,B}^2=\mathbbm{1}$ for any $A$ and $B$, it can be considered the generator of a $\mathbb{Z}_2$ symmetry. $\mathrm{SWAP}_{\Lambda_1,\Lambda_2}$, which exchanges the two spin chains $\Lambda_1$ and $\Lambda_2$, is a $\mathbb{Z}_2$ symmetry of $U_{1,2}$ because it commutes with $U_{1,2}$:
\begin{equation}\label{swapcomm}
U_{1,2}^\dagger \mathrm{SWAP}_{\Lambda_1,\Lambda_2}U_{1,2}=\mathrm{SWAP}_{\Lambda_1,\Lambda_2}.
\end{equation}

\begin{figure}[bt]
   \centering
   \includegraphics[width=.9\columnwidth]{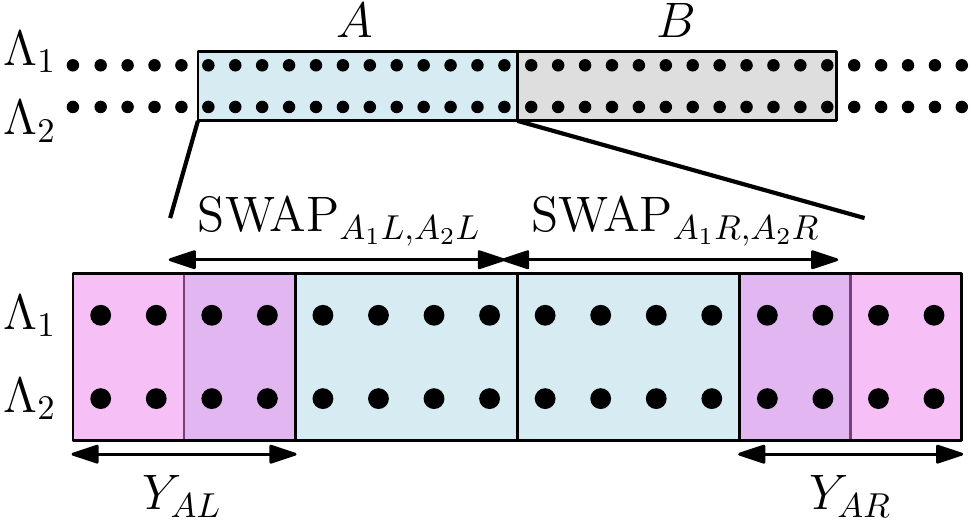} 
   \caption{The GNVW index can be written in terms of SWAP operators in a doubled system $\Lambda_1\cup \Lambda_2$ for two adjacent intervals $A=A_1\cup A_2$ and $B=B_1\cup B_2$. $U_{1,2}=U\otimes U$ modifies $\mathrm{SWAP}_{A_1,A_2}$ by local operators $Y_{AL}$ and $Y_{AR}$ supported near the left and right endpoints of $A$ respectively.}
   \label{fig:swap}
\end{figure}

(\ref{swapcomm}) and the fact that $U_{1,2}$ is locality preserving means that it must act on $\mathrm{SWAP}_{A_1,A_2}$ over some interval $A=A_1\cup A_2$ as
\begin{align}
\begin{split}\label{YLYR}
U_{1,2}^\dagger &(\mathrm{SWAP}_{A_1L,A_2L}\otimes\mathrm{SWAP}_{A_1R,A_2R})U_{1,2}\\
&=Y_{AL}(\mathrm{SWAP}_{A_1L,A_2L}\otimes\mathrm{SWAP}_{A_1R,A_2R})Y_{AR},
\end{split}
\end{align}
where $Y_{AL}$ and $Y_{AR}$ are local operators on the left and right ends of $A$, as shown in Fig.~\ref{fig:swap}. The $L$ and $R$ subscripts indicate the left and right halves of the interval $A$. Writing the GNVW index given in (\ref{gnvw}) using (\ref{swapindex}), and then using (\ref{YLYR}), we obtain

\onecolumngrid
\begin{align}
\begin{split}\label{indexbigswap}
\mathrm{ind}(U)&=\sqrt{\frac{\mathrm{Tr}\left[Y_{AL}(\mathrm{SWAP}_{A_1L,A_2L}\otimes \mathrm{SWAP}_{A_1R,A_2R})Y_{AR}(\mathrm{SWAP}_{B_1L,B_2L}\otimes\mathrm{SWAP}_{B_1R,B_2R})\right]}{\mathrm{Tr}\left[(\mathrm{SWAP}_{A_1L,A_2L}\otimes \mathrm{SWAP}_{A_1R,A_2R})Y_{BL}(\mathrm{SWAP}_{B_1L,B_2L}\otimes\mathrm{SWAP}_{B_1R,B_2R})Y_{BR}\right]}}\\
&=\sqrt{\frac{\mathrm{Tr}(Y_{AL}\mathrm{SWAP}_{A_1L,A_2L})\mathrm{Tr}(\mathrm{SWAP}_{A_1R,A_2R}Y_{AR}\mathrm{SWAP}_{B_1L,B_2L})\mathrm{Tr}(\mathrm{SWAP}_{B_1R,B_2R})}{\mathrm{Tr}(\mathrm{SWAP}_{A_1L,A_2L})\mathrm{Tr}(\mathrm{SWAP}_{A_1R,A_2R}Y_{BL}\mathrm{SWAP}_{B_1L,B_2L})\mathrm{Tr}(\mathrm{SWAP}_{B_1R,B_2R}Y_{BR})}},
\end{split}
\end{align}
\twocolumngrid
where the trace is taken over the same space in the numerator and the denominator. In the second line, we split the trace over tensor products into products of traces.

Let us first use some identities to simplify the middle terms of the numerator and denominator in (\ref{indexbigswap}). Notice that, since $U_{1,2}^\dagger (\mathrm{SWAP}_{A_1,A_2}\otimes\mathrm{SWAP}_{B_1,B_2})U_{1,2}=Y_{AL}(\mathrm{SWAP}_{A_1,A_2}\otimes\mathrm{SWAP}_{B_1,B_2})Y_{BR}$, the following identity holds:
\begin{equation}\label{YAR}
Y_{AR}=Y_{BL}^\dagger.
\end{equation}
This means that
\begin{align}
\begin{split}
&\mathrm{Tr}(\mathrm{SWAP}_{A_1R,A_2R}Y_{AR}\mathrm{SWAP}_{B_1L,B_2L})\\
&=\mathrm{Tr}(\mathrm{SWAP}_{A_1R,A_2R}Y_{BL}^\dagger\mathrm{SWAP}_{B_1L,B_2L}).
\end{split}
\end{align}
Next, using the fact that $\mathrm{SWAP}_{A_1R,A_2R}$ and $\mathrm{SWAP}_{B_1L,B_2L}$ are Hermitian and mutually commute, we obtain
\begin{align}
\begin{split}
&\mathrm{Tr}(\mathrm{SWAP}_{A_1R,A_2R}Y_{AR}\mathrm{SWAP}_{B_1L,B_2L})\\
&=\mathrm{Tr}(\mathrm{SWAP}_{B_1L,B_2L}^\dagger Y_{BL}^\dagger\mathrm{SWAP}_{A_1R,A_2R}^\dagger)\\
&=\mathrm{Tr}(\mathrm{SWAP}_{A_1R,A_2R}Y_{BL}\mathrm{SWAP}_{B_1L,B_2L})^*
\end{split}
\end{align}

It is convenient to write
\begin{align}
\begin{split}
\mathrm{Tr}&(\mathrm{SWAP}_{A_1R,A_2R}Y_{AR}\mathrm{SWAP}_{B_1L,B_2L})\\
&=|\mathrm{Tr}(\mathrm{SWAP}_{A_1R,A_2R}Y_{AR}\mathrm{SWAP}_{B_1L,B_2L})|e^{i\theta}\\
 \mathrm{Tr}&(\mathrm{SWAP}_{A_1R,A_2R}Y_{BL}\mathrm{SWAP}_{B_1L,B_2L})\\
 &=|\mathrm{Tr}(\mathrm{SWAP}_{A_1R,A_2R}Y_{AR}\mathrm{SWAP}_{B_1L,B_2L})|e^{-i\theta},
 \end{split}
 \end{align}
 
 so that
\begin{equation}
\frac{\mathrm{Tr}(\mathrm{SWAP}_{A_1R,A_2R}Y_{AR}\mathrm{SWAP}_{B_1L,B_2L})}{\mathrm{Tr}(\mathrm{SWAP}_{A_1R,A_2R}Y_{BL}\mathrm{SWAP}_{B_1L,B_2L})}=e^{2i\theta}.
\end{equation}

Substituting this into (\ref{indexbigswap}), we obtain
\begin{align}
\begin{split}
&\mathrm{ind}(U)\\
&=\sqrt{\frac{\mathrm{Tr}(Y_{AL}\mathrm{SWAP}_{A_1L,A_2L})\mathrm{Tr}(\mathrm{SWAP}_{B_1R,B_2R})}{\mathrm{Tr}(\mathrm{SWAP}_{A_1L,A_2L})\mathrm{Tr}(\mathrm{SWAP}_{B_1R,B_2R}Y_{BR})}e^{2i\theta}}.
\end{split}
\end{align}
 
To further simplify the above expression, consider $U_{1,2}$ acting on $\mathrm{SWAP}_{A_1,A_2}\otimes\mathrm{SWAP}_{B_1,B_2}$:
\begin{align}
\begin{split}
U_{1,2}^\dagger &(\mathrm{SWAP}_{A_1,A_2}\otimes\mathrm{SWAP}_{B_1,B_2})U_{1,2}\\
&=(Y_{AL}\mathrm{SWAP}_{A_1,A_2})\otimes(\mathrm{SWAP}_{B_1,B_2}Y_{BR}).
\end{split}
\end{align}

Taking the trace of both sides, we obtain
\begin{align}
\begin{split}
&\mathrm{Tr}(\mathrm{SWAP}_{A_1L,A_2L})\mathrm{Tr}(\mathrm{SWAP}_{B_1R,B_2R})\\
&=\mathrm{Tr}(Y_{AL}\mathrm{SWAP}_{A_1L,A_2L})\mathrm{Tr}(\mathrm{SWAP}_{B_1R,B_2R}Y_{BR}),
\end{split}
\end{align}
where we dropped the common factor of $\mathrm{Tr}(\mathrm{SWAP}_{A_1R,A_2R})\mathrm{Tr}(\mathrm{SWAP}_{B_1L,B_2L})$ on both sides. It follows that 

\begin{equation}
\frac{\mathrm{Tr}(Y_{AL}\mathrm{SWAP}_{A_1L,A_2L})}{\mathrm{Tr}(\mathrm{SWAP}_{A_1L,A_2L})}=\frac{\mathrm{Tr}(\mathrm{SWAP}_{B_1R,B_2R})}{\mathrm{Tr}(\mathrm{SWAP}_{B_1R,B_2R}Y_{BR})},
\end{equation}
so we can rewrite the index as

\begin{equation}
\mathrm{ind}(U)=\sqrt{\left(\frac{\mathrm{Tr}(Y_{AL}\mathrm{SWAP}_{A_1L,A_2L})}{\mathrm{Tr}(\mathrm{SWAP}_{A_1L,A_2L})}\right)^2e^{2i\theta}}.
\end{equation}

In order for this quantity to be real, we must have
\begin{equation}
\frac{\mathrm{Tr}(Y_{AL}\mathrm{SWAP}_{A_1L,A_2L})}{\mathrm{Tr}(\mathrm{SWAP}_{A_1L,A_2L})}=\bigg|\frac{\mathrm{Tr}(Y_{AL}\mathrm{SWAP}_{A_1L,A_2L})}{\mathrm{Tr}(\mathrm{SWAP}_{A_1L,A_2L})}\bigg|e^{-i\theta}.
\end{equation}

This is consistent with $Y_{AL}$ and $Y_{AR}$ carrying opposite phases, since we can add a phase $e^{i\theta}$ to $Y_{AR}$ as long as it cancels with the phase we add to $Y_{AL}$, so that
\begin{align}
\begin{split}
\mathrm{Tr}&(Y_{AL}\mathrm{SWAP}_{A_1L,A_2L})\mathrm{Tr}(\mathrm{SWAP}_{A_1R,A_2R}Y_{AR})\\
&=\mathrm{Tr}(\mathrm{SWAP}_{A_1L,A_2L})\mathrm{Tr}(\mathrm{SWAP}_{A_1R,A_2R}).
\end{split}
\end{align}

Putting this together, we obtain
\begin{align}
\begin{split}\label{indexz2}
\mathrm{ind}(U)&=\bigg|\frac{\mathrm{Tr}(Y_{AL}\mathrm{SWAP}_{A_1L,A_2L})}{\mathrm{Tr}(\mathrm{SWAP}_{A_1L,A_2L})}\bigg|\\
&=\bigg|\frac{\mathrm{Tr}(\mathrm{SWAP}_{B_1R,B_2R})}{\mathrm{Tr}(\mathrm{SWAP}_{B_1R,B_2R}Y_{BR})}\bigg|,
\end{split}
\end{align}
which is precisely $\pi_0(U\otimes U)/\pi_1(U\otimes U)$ for the $\mathbb{Z}_2$ symmetry generated by $\mathrm{SWAP}_{\Lambda_1,\Lambda_2}$. Since $\pi_0(U\otimes U)=\mathrm{ind}(U)^2$, we have
\begin{equation}
\mathrm{ind}(U)=\pi_1(U\otimes U).
\end{equation}

In addition, using (\ref{indices}), we have
\begin{equation}
\frac{1}{\mathrm{ind}(U)}=\left|\frac{\mathrm{Tr}(R_1)}{\mathrm{Tr}(U_{1,R})}\right|,
\end{equation}
where $R_1$ is defined for $U_{1,2}$ acting on the doubled system (not $U$).

As a sanity check, suppose that $U$ acts as a translation by a single site to the right on a chain of $d$-state spins. Then $Y_{AL}$ is a $\mathrm{SWAP}$ operator on the leftmost site in $A$ and
\begin{equation}
\frac{\mathrm{Tr}(Y_{AL}\mathrm{SWAP}_{A_1L,A_2L})}{\mathrm{Tr}(\mathrm{SWAP}_{A_1L,A_2L})}=\frac{d^{2}d^{N_{AL}-1}}{d^{N_{AL}}}=d,
\end{equation}
where $N_{AL}=\frac{1}{2}|A_1|=\frac{1}{2}|A_2|$. Alternatively, $Y_{BR}$ is a $\mathrm{SWAP}$ operator on one additional site to the right of $B_R$, so

\begin{equation}
\frac{\mathrm{Tr}(\mathrm{SWAP}_{B_1R,B_2R})}{\mathrm{Tr}(\mathrm{SWAP}_{B_1R,B_2R}Y_{BR})}=\frac{d^{N_{BR}}d^2}{d^{N_{BR}+1}}=d,
\end{equation}
as expected. 


%
%
%
%
%

Note that the above derivation suggests a way to measure $\pi_{1}(U)$ for a general $\mathbb{Z}_2$ symmetric QCA:

\begin{equation}
\frac{\pi_0(U)}{\pi_{1}(U)}=\sqrt{\frac{\mathrm{Tr}(U^\dagger U_{1,A}U U_{1,B})}{\mathrm{Tr}(U_{1,A}U^\dagger U_{1,B}U)}},
\end{equation}
where $A$ and $B$ are adjacent intervals and $U_{1,A}$ and $U_{1,B}$ are the symmetry actions for the nontrivial group element on these intervals. 

\subsection{Mutual information formulation of the GNVW index}\label{snosymm}
\begin{figure}[tb]
   \centering
   \includegraphics[width=.9\columnwidth]{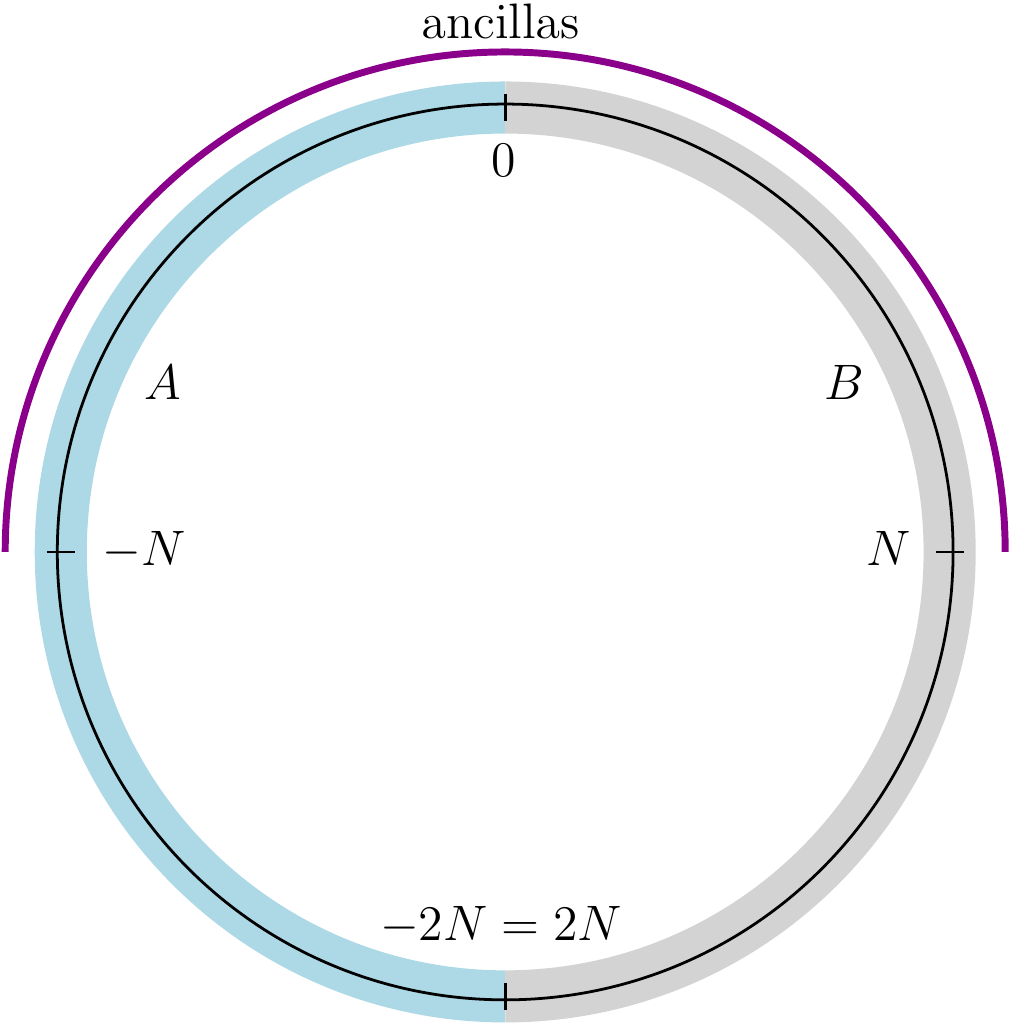} 
   \caption{The setup for the mutual information formula for the GNVW index. We consider a periodic spin chain with sites labeled by $[-2N+1,2N]$ with site $-2N$ identified with site $N$. We label $A=[-2N+1,0]$ (blue) and $B=[1,2N]$ (grey). Each site in $[-N+1,N]$ is initially maximally entangled with a partner ancilla (purple).}
   \label{fig:info}
\end{figure}

We will now show that (\ref{indexz2}) and a second Renyi version of the mutual information formula for the GNVW index given in Ref.~\onlinecite{tracking} (see also Ref.~\onlinecite{ranard2022}) are equal. 

The setup in Ref.~\onlinecite{tracking}, as shown in Fig.~\ref{fig:info}, consists of $U$ acting on a periodic spin chain whose sites we label $[-2N+1,2N]$, with site $-2N$ identified with site $2N$. Note that Ref.~\onlinecite{tracking} considered a 2D disk evolving under a Floquet circuit, but in this work we will only consider a 1D spin chain consisting of the spins near the edge of the disk, evolving under the effective edge unitary of the 2D Floquet circuit. We label the left half of the spin chain as $A=[-2N+1,0]$ and the right half of the spin chain as $B=[1,2N]$. In addition, we include a chain of ancillas with the same local Hilbert space as the original sites, each maximally entangled with a state on spin chain, on the interval $[-N+1,N]$. The density matrix for the entire system including the ancillas has the form
\begin{equation}
\rho=\rho_{[-2N+1,-N]}\otimes\rho_{[-N+1,N]}^a\otimes\rho_{[N+1,2N]},
\end{equation}
where $\rho_{[-2N+1,-N]}$ and $\rho_{[N+1,2N]}$ are density matrices of the spins in the spin chain in a product state and $\rho_{[-N+1,N]}^a$ is the density matrix describing each spin on the chain maximally entangled with its partner ancilla. The action of $U$ on $\rho$ gives

\begin{equation}
U^\dagger \rho U=\tilde{\rho}.x
\end{equation}

Note that we evolve the density matrix using $U^\dagger \rho U$ rather than with $U$ and $U^\dagger$ switched, because we use the convention that a translation in the positive direction moves \emph{operators} in the positive direction and \emph{states} in the opposite direction. This is also the convention used in Refs.~\onlinecite{u1floquet,flows}.

In this setup, Ref.~\onlinecite{tracking} proposed the following equation for $\log\mathrm{ind}(U)$:
\begin{align}
\begin{split}\label{mutinf}
\nu&=\log\mathrm{ind}(U)\\
&=\mathcal{I}(a_A,\rho_B)-\mathcal{I}(a_B,\rho_A).
\end{split}
\end{align}

Here, $a_A(a_B)$ are the reduced density matrices of the ancillas associated with sites $[-N+1,0]$ $\left([1,N]\right)$ and $\rho_A$ $\left(\rho_B\right)$ are the reduced density matrices of the spins on sites $[-2N+1,0]$ $\left([1,2N]\right)$. Specifically, $a_A$ is the reduced density matrix obtained by starting with $\tilde{\rho}$ and tracing out all the spins and ancillas other than the ancillas paired with spins on sites $[-N+1,0]$. The mutual information $\mathcal{I}(A,B)$ between any two sets of sites $A$ and $B$ is defined as 
\begin{equation}
\mathcal{I}(A,B)=\frac{1}{2}\left[S(\rho_A)+S(\rho_B)-S(\rho_{A\cup B})\right].
\end{equation}

Using the complementarity property of entanglement, (\ref{mutinf}) simplifies to\cite{tracking}
\begin{equation}\label{entindex}
\nu=\frac{1}{2}\left[S(\rho_B)-S(\rho_A)\right].
\end{equation}

We will now relate a second Renyi version of this quantity to (\ref{indexz2}). We will use the second Renyi entropy because it can be conveniently written using a SWAP operator on a doubled system\cite{hastings2010}:
\begin{align}
\begin{split}
S_2(\rho_A)&=-\log\mathrm{Tr}(\rho_A^2)\\
&=-\log\mathrm{Tr}\left(\mathrm{SWAP}_{A_1,A_2}(\rho\otimes\rho)\right),
\end{split}
\end{align}
where $\mathrm{SWAP}_{A_1,A_2}$ acts on two identical copies of the original system. In terms of Renyi entropies, (\ref{entindex}) reads
\begin{equation}\label{entanglement}
\nu=\frac{1}{2}\log\frac{\mathrm{Tr}\left(\mathrm{SWAP}_{A_1,A_2}(\tilde{\rho}\otimes\tilde{\rho})\right)}{\mathrm{Tr}\left(\mathrm{SWAP}_{B_1,B_2}(\tilde{\rho}\otimes\tilde{\rho})\right)}.
\end{equation}

Using cyclicity of the trace, we obtain
\begin{align}
\begin{split}
\mathrm{Tr}&(\mathrm{SWAP}_{A_1,A_1}(\tilde{\rho}\otimes\tilde{\rho}))\\
&=\mathrm{Tr}\left(U_{1,2}\mathrm{SWAP}_{A_1,A_2}U_{1,2}^\dagger (\rho\otimes\rho)\right).
\end{split}
\end{align}

$\nu$ is additive under composition of QCA, so $\nu\to-\nu$ for $U\to U^\dagger$. Therefore, we can write $\mathrm{ind}(U)$ from (\ref{entanglement}) as
\begin{equation}\label{entanglement2}
\mathrm{ind}(U)=\sqrt{\frac{\mathrm{Tr}\left(U_{1,2}^\dagger \mathrm{SWAP}_{B_1,B_2}U_{1,2}(\rho\otimes\rho)\right)}{\mathrm{Tr}\left(U_{1,2}^\dagger \mathrm{SWAP}_{A_1,A_2}U_{1,2}(\rho\otimes\rho)\right)}}.
\end{equation}

Now we can use (\ref{YLYR}) to obtain
\begin{align}
\begin{split}
\mathrm{Tr}&\left(U_{1,2}^\dagger \mathrm{SWAP}_{B_1,B_2}U_{1,2}(\rho\otimes\rho)\right)\\
&=\mathrm{Tr}\left(Y_{BL}\mathrm{SWAP}_{B_1,B_2}Y_{BR}(\rho\otimes\rho)\right)\\
&=\mathrm{Tr}\left(Y_{BL}\mathrm{SWAP}_{B_1L,B_2L}\rho_{[-N+1,N]}\right)\\
&\times \mathrm{Tr}\left(\mathrm{SWAP}_{B_1R,B_2R}Y_{BR}\rho_{[N+1,-N]}\right).
\end{split}
\end{align}
In the last line, we factorized the trace and use $\rho_{[N+1,-N]}$ as shorthand for $\rho_{[N+1,2N]}\rho_{[-2N+1,-N]}$. The SWAP operators, as well as $Y_{BL}$ and $Y_{BR}$, do not act on the ancillas. Therefore, we can easily trace over the ancillas to get
\begin{align}
\begin{split}
\mathrm{Tr}&\left(U_{1,2}^\dagger \mathrm{SWAP}_{B_1,B_2}U_{1,2}(\rho\otimes\rho)\right)\\
&=\frac{1}{d^{4N}}\mathrm{Tr}\left(Y_{BL}\mathrm{SWAP}_{B_1L,B_2L}\right)\\
&\times \mathrm{Tr}\left(\mathrm{SWAP}_{B_1R,B_2R}Y_{BR}\rho_{[N+1,-N]}\right).
\end{split}
\end{align}

We can apply similar manipulations to the denominator of (\ref{entanglement2}). Then, using the fact that $\rho_{[N+1,-N]}$ is a product state density matrix and $\mathrm{Tr}(Y_{BR}\rho_{[N+1,-N]})=\mathrm{Tr}(Y_{AL}\rho_{[N+1,-N]})^*$, we get
\begin{equation}\label{indmatch}
\mathrm{ind}(U)=\sqrt{\left|\frac{\mathrm{Tr}(Y_{BL}\mathrm{SWAP}_{B_1LB_2L})}{\mathrm{Tr}(\mathrm{SWAP}_{A_1R,A_2R}Y_{AR})}\right|}.
\end{equation}

This is equal to GNVW index as written in (\ref{indexz2}). Specifically, (\ref{indmatch}) is obtained by taking the square root of the product of the first and second lines of (\ref{indexz2}), then using $|A|=|B|$ and a relabeling of $A$ and $B$.

\section{Discussion}\label{sdiscussion}
In this note, we have presented a few results on the classification of QCA with symmetry, using strong equivalence. We rederived the classification given in Ref.~\onlinecite{mpu} without the extra framework of tensor networks. Furthermore, we showed that the refined SPI of Ref.~\onlinecite{mpu} are only complete for certain symmetry groups. For the particular case of $G=\mathbb{Z}_2$, we were able to obtain a complete classification, and furthermore show that the GNVW index can be interpreted in terms of a $\mathbb{Z}_2$ SPI in a doubled system. Using this SPI formulation of the GNVW index, we derived a second Renyi version of the mutual information formulation of the index studied in Ref.~\onlinecite{tracking}. 

A number of interesting questions remain. First, we still do not have a concrete formula for invariants that completely classify QCA with discrete symmetry, when the group is not $\mathbb{Z}_N$ for $N$ prime. The main difficulty is finding easily computable quantities that remove the ambiguity of the 1D representation attached to $R_g$.

Second, we showed that a second Renyi version of the mutual information formula of Ref.~\onlinecite{tracking} matches with the GNVW index if we start in a state where the ancillas are maximally entangled with their respective spin in the spin chain. However, Ref.~\onlinecite{tracking} showed numerically that the mutual information formula still well approximates the GNVW index even when the ancillas are \emph{not} maximally entangled with their respective spin. A deeper study of generalizations of the derivation presented in this work may help understand this intriguing robustness of the mutual information calculation.

\acknowledgments

We thank Michael Levin and Zongping Gong for helpful discussions and for comments on the draft. We especially thank Benjamin Krakoff for collaboration on the results presented in Sec.~\ref{slim}.
This work was supported by the National Science Foundation Graduate Research Fellowship under Grant No. 1746045 and the University of Chicago Bloomenthal Fellowship.
\appendix
\section{Derivation of the Classification}\label{sderivation}

In this appendix, we will show that 1D QCA with $G$ symmetry are equivalent if and only if their corresponding representations $R_g$ as defined in (\ref{rgdef}) differ only by a 1D representation $u_g$ and tensoring with copies of the on-site representation $\mu_g$. Due to the stacking structure of the QCA, we only need to prove that $W$ is a $G$ symmetric FDQC if and only if $R_g=u_g\mu_g^{\otimes N}$ for some nonnegative integer $N$. The methods used here closely follow those used in Ref.~\onlinecite{u1floquet}. 

\begin{figure}[tb]
   \centering
   \includegraphics[scale=1.]{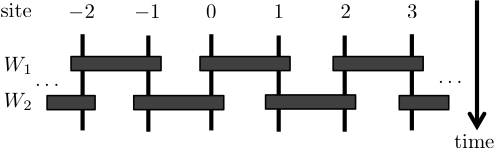} 
   \caption{$W$ written as a depth two unitary of layers of commuting two-(super)site gates.}
   \label{fig:depth2}
\end{figure}

First, we will show that if $R_g=u_gU_{g,R}=u_g\mu_g^{\otimes N}$, then $W$ is a $G$ symmetric FDQC. Since $\mathrm{dim}(R_g)=\mathrm{dim}(u_gU_{g,R})$, the GNVW index must be trivial. Therefore, without loss of generality, we assume that $W$ is a depth two circuit where each layer consists of disjoint local unitaries acting over a distance $\xi$. We now cluster together neighboring sites into supersites such that each gate of the FDQC acts on two neighboring supersites, and set $\xi=1$ in units of the supersites. The unitary representation of $G$ for a supersite is simply $\mu_g^{\otimes N}=U_{g,R}$ where $N$ is the number of original sites per supersite. We will denote $U_{g,R}$ by $\rho_g$ for brevity, and we denote the symmetry action on site $i$ by $\rho_g^{(i)}$. $W$ is given by 

\begin{align}
W = W_2W_1= \prod_i W_2^{(2i-1,2i)}\prod_i W_1^{(2i,2i+1)} .
\label{w1w2}
\end{align}
This decomposition of $W$ is illustrated in Fig \ref{fig:depth2}. 

Suppose we act $W$ on $\rho_g^{(0)}\otimes \rho_g^{(1)}$. Because $W$ is an LPU and $W$ commutes with the symmetry operators for the entire system $U_g$, we have

\begin{equation}
W(\rho_g^{(0)}\otimes\rho_g^{(1)})W^\dagger=(Y_{g,L}\rho_g^{(0)})\otimes(\rho_g^{(1)}Y_{g,R}),
\end{equation}
where $Y_{g,L}$ and $Y_{g,R}$ are operators supported on supersites $[-1,0]$ and $[1,2]$ respectively. According to the notation of (\ref{rgdef}), $Y_{g,L}\rho_g^{(0)}=L_g$ and $\rho_g^{(1)}Y_{g,R}=R_g$.  

Notice that there is a phase ambiguity in the definition of $Y_{g,L}$ and $Y_{g,R}$. While $Y_{g,L}\otimes Y_{g,R}$ forms a linear representation of $G$, $Y_{g,L}$ and $Y_{g,R}$ individually form projective representations. We consider only trivial projective representations to focus on strong equivalence phases. Specifically, in the case of trivial projective representations, we can lift $Y_{g,R}$ to a linear representation by relabeling $Y_{g,R}\to e^{-i\theta_g}Y_{g,R}$ and similarly relabeling $Y_{g,L}$, where $e^{-i\theta_g}\in U(1)$. The result however still has an ambiguity associated with multiplying $Y_{g,R}$ with a 1D representation of $G$, denoted $u_g$. We will revisit this ambiguity shortly.

Using the form of $W$ in Eq.~\ref{w1w2}, we obtain
\begin{equation}
W_1\left(\rho_g^{(0)}\otimes\rho_g^{(1)}\right)W_1^\dagger=W_2^\dagger\left(Y_{g,L}\rho_g^{(0)}\otimes \rho_g^{(1)}Y_{g,R}\right)W_2.
\end{equation}

Writing $W_1$ and $W_2$ in terms of two-site gates gives:
\begin{align}
\begin{split}\label{Wtilderho}
&W_1^{(0,1)}\left(\rho_g^{(0)}\otimes\rho_g^{(1)}\right)W_1^{(0,1)\dagger}\\
&=W_2^{(-1,0)\dagger}W_2^{(1,2)\dagger}\left(Y_{g,L}\rho_g^{(0)}\otimes \rho_g^{(1)}Y_{g,R}\right)W_2^{(1,2)}W_2^{(-1,0)}\\
&=W_2^{(-1,0)\dagger}(Y_{g,L}\rho_g^{(0)})W_2^{(-1,0)}\otimes W_2^{(1,2)\dagger}(\rho_g^{(1)}Y_{g,R})W_2^{(1,2)}.
\end{split}
\end{align}

Substituting
\begin{align}
\begin{split}
\tilde{\rho}_g^{(0)}&=W_2^{(-1,0)\dagger}(Y_{g,L}\rho_g^{(0)})W_2^{(-1,0)}\\
\tilde{\rho}_g^{(1)}&=W_2^{(1,2)\dagger}(\rho_g^{(1)}Y_{g,R})W_2^{(1,2)}
\end{split}
\end{align}
gives
\begin{equation}
W_1^{(0,1)}\left(\rho_g^{(0)}\otimes\rho_g^{(1)}\right)W_1^{(0,1)\dagger}=\tilde{\rho}_g^{(0)}\otimes \tilde{\rho}_g^{(1)}.
\end{equation}

The left side is supported only on sites $[0,1]$ and the right side has two terms, one supported on $[-1,0]$ and one supported on $[1,2]$. $W$ being trivial under strong equivalence means that it can be made $G$ symmetric. This means that there exists $\tilde{W}_1$ and $\tilde{W}_2$ that are individually $G$ symmetric. This is the case if and only if there are on-site unitaries $V_i$ such that

\begin{equation}\label{Viunitary}
V_0\tilde{\rho}_g^{(0)}V_0^\dagger =\rho_g^{(0)}\qquad V_1 \tilde{\rho}_g^{(1)}V_0^\dagger=\rho_g^{(1)}.
\end{equation}

Then we can define
\begin{equation}
\tilde{W}_2^{(i,j)}=W_2^{(i,j)}V_j^\dagger V_i^\dagger \qquad \tilde{W}_1^{(i,j)}= V_iV_jW_1^{(i,j)}.
\end{equation}

We can see that $\tilde{W}_1$ is $G$-symmetric:
\begin{align}
\begin{split}
V_1 &V_0 W_1^{(0,1)}\left(\rho_g^{(0)}\otimes\rho_g^{(1)}\right)W_1^{(0,1)\dagger}V_0^\dagger V_1^\dagger \\
&=V_1 V_0\left(\tilde{\rho}_g^{(0)}\otimes \tilde{\rho}_g^{(1)}\right)V_0^\dagger V_1^\dagger \\
&=\rho_g^{(0)}\otimes\rho_g^{(1)},
\end{split}
\end{align}
and since $W$ is $G$-symmetric and $W=W_2W_1=\tilde{W}_2\tilde{W}_1$, $\tilde{W}_2$ is also $G$-symmetric.

Notice that as long as $\rho_g^{(0)}$ and $Y_{g,L}\rho_g^{(0)}$ have the same eigenvalues up to multiplication by a 1D representation of $G$ (as do $\rho_g^{(1)}$ and $\rho_g^{(1)}Y_{g,R}$), there exists such on-site unitaries $V_i$ with action given by (\ref{Viunitary}). In this case,
\begin{equation}
\mathrm{Tr}(\rho_gY_{g,R})=u_g\mathrm{Tr}(\rho_g)\qquad \forall g\in G,
\end{equation}
where $u_g$ is a 1D representation of $G$. This also means that the representation $\rho_g$ and the representation $\rho_gY_{g,R}=R_g$ must be equivalent modulo multiplication by 1D representations of $G$. Since $\rho_g=\mu_g^{\otimes N}$, we obtain the desired result.

We now show that if $W$ is a $G$ symmetric FDQC (with the addition of ancillas carrying the $\mu_g$ representation of $G$), then $R_g=u_g\rho_g^{\otimes N}$ for some nonnegative integer $N$. Since $W$ is a FDQC, for an interval $A$ larger than $2\xi$, we can write
\begin{equation}
W=W_AW_{A^c}W_LW_R,
\end{equation}
where $W_A$ is supported fully in $A$, $W_{A^c}$ is supported fully outside of $A$, and $W_L$ and $W_R$ are supported near the left and right endpoints of $A$ respectively. The fact that $W$ is a $G$ symmetric FDQC means that the above four unitary operators all $G$ symmetric. It follows that
\begin{equation}\label{waeq}
[W_A,U_{g,L}U_{g,R}]=0.
\end{equation}
Furthermore, because $|A|>2\xi$, we see that due to disjoint operator support, we have
\begin{equation}\label{wlwr}
[W_L,U_{g,R}]=[W_R,U_{g,L}]=[W_{A^c},U_{g,L}U_{g,R}]=0.
\end{equation}

Using (\ref{waeq}) and (\ref{wlwr}), we have
\begin{align}
\begin{split}
W^\dagger \left(U_{g,L}U_{g,R}\right) W &=W_R^\dagger W_L^\dagger\left(U_{g,L}U_{g,R}\right)W_LW_R\\
&=\left(W_L^\dagger U_{g,L}W_L\right)\left(W_R^\dagger U_{g,R}W_R\right).
\end{split}
\end{align}

Adding ancillas carrying the same representation of $G$ corresponds to tensoring copies of $\mu_g$ to $U_{g,L}$ and $U_{g,R}$ and adding symmetric gates to $W$ to couple the ancillas with the original spins. Denote the representation of the symmetry with the ancillas by $U_{g,R}'$ and $W$ with the ancillas by $W'=W_A'W_{A^c}'W_L'W_R'$. Since $\left(W_R^\dagger U_{g,R}W_R\right)$ and $R_g$ are both supported to the right of the midpoint of $A$, we have
\begin{equation}
R_g=u_gW_R^{\dagger'} U_{g,R}'W_R'
\end{equation}

Similarly, we have
\begin{equation}
L_g=u_g^\dagger W_L^{\dagger'} U_{g,L}'W_L'
\end{equation}
where again, $u_g$ is a 1D representation of $G$. Since $W_R'$ and $W_L'$ are unitary operators, they cannot change the spectrum of $U_{g,R}'$ and $U_{g,L}'$ respectively. It follows that $R_g=u_g\mu_g^{\otimes N}$ and $L_g=u_g^\dagger\mu_g^{\otimes N}$ for some nonnegative integer $N$, as desired. Notice that if we allowed the addition of ancillas carrying any representation of the symmetry, then we can tensor any representation to $U_{g,L}$ and $U_{g,R}$, so $R_g$ can be any linear representation of $G$.


%
%
%

\bibliography{flowsfloquetbib}

\end{document}